\documentclass[twocolumn,prl,aps,superscriptaddress]{revtex4-1}
\usepackage[latin9]{inputenc}
\usepackage{mathtools}
\usepackage{amsbsy}
\usepackage{amstext}
\usepackage[unicode=true,pdfusetitle,
 bookmarks=false,
 breaklinks=false,pdfborder={0 0 1},backref=false,colorlinks=false]
 {hyperref}
\usepackage[dvipsnames]{xcolor}
 \hypersetup{colorlinks=true,citecolor=NavyBlue,filecolor=black,linkcolor=black,urlcolor=black}

\makeatletter

\@ifundefined{textcolor}{}
{%
 \definecolor{BLACK}{gray}{0}
 \definecolor{WHITE}{gray}{1}
 \definecolor{RED}{rgb}{1,0,0}
 \definecolor{GREEN}{rgb}{0,1,0}
 \definecolor{BLUE}{rgb}{0,0,1}
 \definecolor{CYAN}{cmyk}{1,0,0,0}
 \definecolor{MAGENTA}{cmyk}{0,1,0,0}
 \definecolor{YELLOW}{cmyk}{0,0,1,0}
}

\newcommand\ket[1]{\left|#1\right\rangle}
\newcommand\bra[1]{\left\langle #1 \right|}

\usepackage{amsmath}
\usepackage{graphicx}
\usepackage{amssymb}
\usepackage{txfonts,color}
\usepackage{dsfont}
\makeatother

\begin{document}
\title{Harnessing non-adiabatic excitations promoted by a quantum critical point}
\author{Obinna Abah}
\affiliation{Centre for Theoretical Atomic, Molecular and Optical Physics, Queen's University Belfast, Belfast BT7 1NN, United Kingdom}
\author{Gabriele De Chiara}
\affiliation{Centre for Theoretical Atomic, Molecular and Optical Physics, Queen's University Belfast, Belfast BT7 1NN, United Kingdom}
\author{Mauro Paternostro}
\affiliation{Centre for Theoretical Atomic, Molecular and Optical Physics, Queen's University Belfast, Belfast BT7 1NN, United Kingdom}
\author{Ricardo Puebla}
\affiliation{Instituto de F{\'i}sica Fundamental, IFF-CSIC, Calle Serrano 113b, 28006 Madrid, Spain}
\affiliation{Centre for Theoretical Atomic, Molecular and Optical Physics, Queen's University Belfast, Belfast BT7 1NN, United Kingdom}

\begin{abstract}
Crossing a quantum critical point in finite time challenges the adiabatic condition due to the closing of the energy gap, which ultimately results in the formation of excitations. Such non-adiabatic excitations are typically deemed detrimental in many scenarios, and consequently several strategies have been put forward to circumvent their formation. Here, however, we show how these non-adiabatic excitations -- originated from the failure to meet the adiabatic condition due to the presence of a quantum critical point -- can be controlled and thus harnessed to perform certain tasks advantageously. We focus on closed cycles reaching the quantum critical point of fully-connected models analyzing two examples. First, a quantum battery that is loaded by approaching a quantum critical point, whose stored and extractable work increases exponentially via repeating cycles. Second, a scheme for the fast preparation of spin squeezed states containing multipartite entanglement that offer a metrological advantage. The corresponding figure of merit in both cases crucially depends on universal critical exponents and the scaling of the protocol driving the system in the vicinity of the transition. Our results highlight the rich interplay between quantum thermodynamics and metrology with critical nonequilibrium dynamics.
\end{abstract}

\maketitle
{\em Introduction.---} Nonequilibrium dynamics triggered by a quantum phase transition (QPT) is one of the most fascinating aspects in the area of quantum many-body systems~\cite{Polkovnikov:11,Eisert:15}. The understanding of nonequilibrium scaling behavior near a quantum critical point has attracted significant attention in condensed matter and statistical physics~\cite{Sachdev,Polkovnikov:11,Eisert:15,delCampo:14}. The ongoing theoretical efforts are benefiting from the extraordinary experimental progress made during the last decades in ultracold atomic and molecular gases, trapped ions and solid state systems, where quantum many-body systems can now be realized and controlled to a very high degree of isolation~\cite{Bloch:08,Weiler:08,Kim:10,Blatt:12,Langen:15,Zhang:17b,Jurcevic:17,Wilkinson:20}. These advances are opening exciting new directions aiming at exploiting and harnessing quantum effects to perform different tasks~\cite{Dowling:03,Deutsch:20}, for instance in the fields of quantum computation and information processing~\cite{Nielsen,Preskill:18}, quantum sensing~\cite{Degen:17,Pezze:18} and quantum thermodynamics~\cite{Deffner,Binder}. 
In this regard, a finite-time modulation of a control parameter in a quantum system may be essential to carry out the desired goal as, for example, in quantum thermodynamic engines~\cite{Andresen:2011,Abah:19} or in quantum annealing and adiabatic quantum computation~\cite{Das:08}. However, a finite-time evolution may entail non-adiabatic excitations, which are typically detrimental to the performance and, therefore, different strategies have been proposed to circumvent their formation~\cite{Werschnik:07,GueryOdelin:19,Power:13,GomezRuiz:19,Wu:15}. This is particularly relevant when the evolution traverses or reaches a QPT, since the vanishing energy gap at the critical point leads to the breakdown of the adiabatic condition~\cite{Polkovnikov:08}, and the formation of excitations, as described within the Kibble-Zurek mechanism~\cite{Zurek:05,Polkovnikov:05,Dziarmaga:05,delCampo:14}.

Among the distinct directions within the flourishing field of quantum thermodynamics~\cite{Deffner,Binder}, quantum many-body systems acting as working medium of heat engines or as quantum batteries are attracting considerable attention in the quest to enhance thermodynamic performance thanks to quantum effects~\cite{Binder:15,Campisi:16,Ma:17,Thao:18,Ferraro:18,Andolina:19,Andolina:19b,Rossini:19,Crescente:20,Revathy:20,Liu:21}. Yet, the resulting non-adiabatic excitations due to existence of a critical point can impair their performance~\cite{Fogarty:20}. 
Hence, underpinning scenarios where the existence of a QPT can be advantageous is of primarily importance to devise strategies to scale up quantum machines and disclose novel settings to achieve quantum enhancement. 


Conversely, it is well-known that the ground state at a QPT can serve as a resource for quantum information processing tasks~\cite{Amico:08,Luo:17,Chiara:18} due their large degree of entanglement, which can also be beneficial in quantum metrology~\cite{Zanardi:08,Pezze:09,Rams:18,Garbe:20,Chu:21,Liu:21} and quantum thermometry~\cite{Mehboudi:19}. Nevertheless, the preparation of such critical ground states is challenged by the breakdown of the adiabatic condition, and thus either long evolution times and/or sophisticated protocols are typically required to exploit their critical features. In this manner, new strategies capable of yielding resourceful quantum states while minimizing the time and complex protocols are very desirable.

In this Letter, we tackle these problems and show that non-adiabatic excitations, formed as a consequence of the vanishing energy gap at a critical point, can be harnessed and controlled in critical fully-connected models featuring a QPT~\footnote{The investigation of spatially-extended systems such as interacting spin chains featuring a QPT is left for a future work.}, whose effective description reduces to a driven quantum harmonic oscillator, such as the quantum Rabi model~\cite{Hwang:15,Puebla:16,Puebla:20} or Lipkin-Meshkov-Glick (LMG) model~\cite{Lipkin:65,Ribeiro:07,Ribeiro:08}, among others.
We consider closed cycles in these systems reaching the critical point, and show the advantage that such non-adiabatic excitations 
entail for two different scenarios. 

On the one hand, promoting excitations in a critical system can be considered as a resource to produce and store work. Besides a full characterization of the accumulated work and fluctuations thereof, we demonstrate that the consecutive repetition of $M$ cycles can lead to an exponential increase of the stored and extractable work from such quantum critical battery. On the other hand, we show how to harness these non-adiabatic excitations to yield a large degree of spin squeezing in the LMG model, which describes the long-range interaction of $N$ spin-$1/2$ particles. We show that such spin squeezed states contain multipartite entanglement and are therefore useful for quantum metrology tasks~\cite{Pezze:18,Pezze:09}. Owing to the inherently non-adiabatic nature of the protocol, such states are obtained in a fast fashion. Our results, which can be readily applied to different experimental platforms, highlight the rich interplay between quantum thermodynamics, quantum metrology and critical nonequilibrium dynamics.

\begin{figure}
\centering
\includegraphics[width=0.8\linewidth,angle=-0]{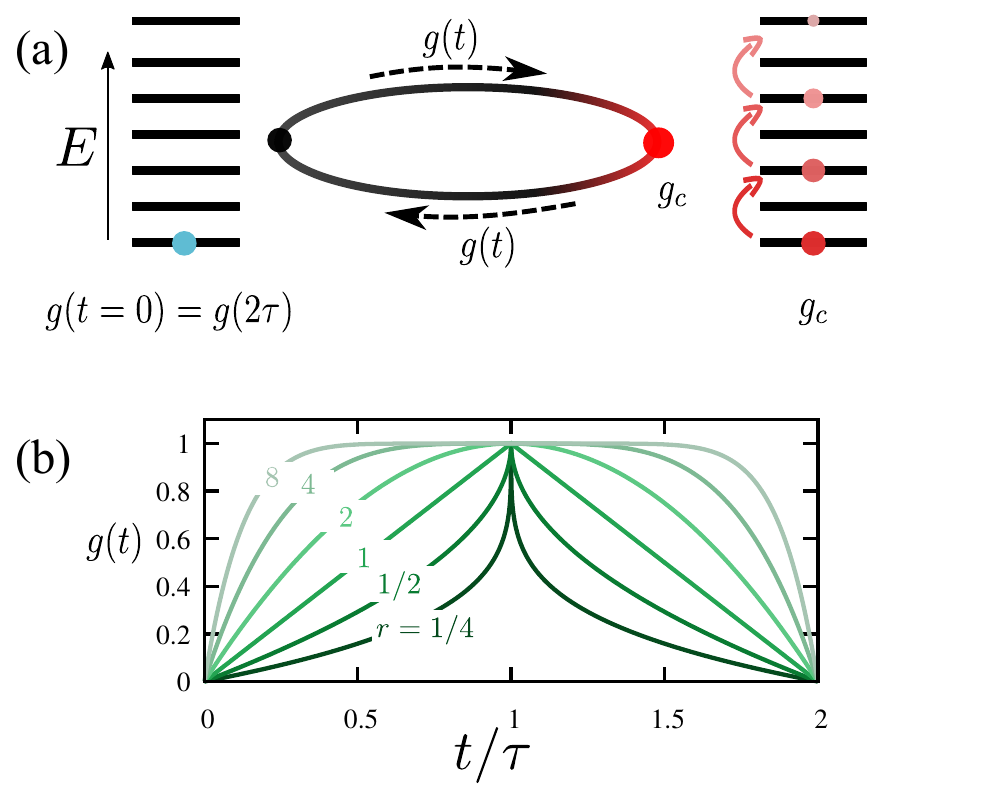}
\caption{\small{(a) Sketch of the cyclic transformation to harness non-adiabatic excitations formed in the vicinity of the critical point $g_c$ due to the closing of the energy gap. In light of the Hamiltonian in Eq.~\eqref{eq:H0}, the state becomes squeezed. (b) Illustration of the different profiles of $g(t)$ for different nonlinear exponents $r$ (cf. Eq.~\eqref{eq:gt}).}}
\label{fig1}
  \end{figure}

{\em Preliminaries.---} We shall start our analysis considering a driven quantum harmonic oscillator ($\hbar=1$),
\begin{align}\label{eq:H0}
    H(t)=\omega a^\dagger a-\frac{g^2(t)\omega}{4}(a+a^\dagger)^2
\end{align}
 which effectively captures the critical features of different fully-connected models such the LMG model~\cite{Lipkin:65,Ribeiro:07,Ribeiro:08}, the critical quantum Rabi model and related systems displaying a superradiant phase transition~\cite{Dicke:54,Emary:03,Emary:03prl,Hwang:15,Puebla:16,Peng:19,Zhu:20}, as well as different realizations of Bose-Einstein condensates~\cite{Anquez:16,Mottl:12,Brennecke:13,Zibold:10}. It is worth mentioning that these models have been realized experimentally~\cite{Anquez:16,Mottl:12,Brennecke:13,Zibold:10,Jurcevic:17,Cai:21}. The Hamiltonian of Eq.~\eqref{eq:H0} is a valid description of these models for $|g|\leq g_c=1$, where $g_c$ denotes the critical point at which the energy gap $\epsilon(g) =\omega\sqrt{1-g^2}$ vanishes as $\epsilon(g)\approx |g-g_c|^{z\nu}$ with $z\nu=1/2$ dependent on the critical exponents of the QPT~\cite{Sachdev}. The harmonic oscillator is described in terms of the standard creation and annihilation operators, $[a,a^\dagger]=1$, and is driven according to a cyclic transformation reaching the critical point $g_c$. In particular, the time-dependent protocol reads as
 \begin{align}\label{eq:gt}
     g(t)=\begin{cases}g_c\left(1-\frac{(\tau-t)^r}{\tau^r}\right)\quad 0\leq t \leq \tau\\
     g_c\left(1-\frac{(t-\tau)^r}{\tau^r} \right) \quad \tau < t \leq 2\tau
     \end{cases}
 \end{align}
 where we have assumed, without loss of generality, that $g(0)=g(2\tau)=0$. The nonlinear exponent $r>0$ controls how the system approaches the critical region~\cite{Barankov:08}, namely, $|g_c-g(t)|\propto |t-\tau|^r$, and the rate at which the system is driven, $|\dot{g}(t)|=2g_c r|t-\tau|^{r-1}\tau^{-r}$ (cf. Fig.~\ref{fig1}). 
 
 Under these general considerations, one can show that the evolved state of the system at time $t$ reads $\ket{\psi(t)}\propto e^{b(t)(a^{\dagger})^2}\ket{0}$ where $\ket{0}$ is the initial state and the parameter $b(t)$ changes according to the equation of motion~\cite{sup}
 \begin{align}\label{eq:bt}
       \dot{b}(t)=-i\omega\left(2b(t)-\frac{g^2(t)}{4}(1+2b(t))^2\right),
 \end{align}
 with $b(0)=0$. 
 Hence, the state $\ket{\psi(t)}$ is simply a squeezed state (up to an irrelevant global phase),  $\ket{\psi(t)}=S(s)\ket{0}$ with $|s|={\rm artanh}(2|b(t)|)$ and $S(s)=e^{s(a^{\dagger})^2/2-s^*a^2/2}$ the squeezing operator~\cite{sup,Fisher:84,Truax:85,Scully}. By setting $\dot{b}(t)=0$ one recovers the ground state squeezing of Eq.~\eqref{eq:H0}, that is, $|s|=-\log(1-g^2)/4$~\cite{Hwang:15}. 
  
 As a consequence of the breakdown of the adiabatic condition due to a vanishing energy gap $\epsilon(g_c)=0$~\cite{Polkovnikov:08}, the initial state is never retrieved upon a slow cycle, regardless of how slow it is performed, that is, independently of how large  $\omega\tau$ is~\cite{Defenu:20,sup}. In particular, the resulting final state after one cycle for $\omega\tau\gtrsim 1$ acquires a squeezing~\cite{sup}
 \begin{align}\label{eq:s}
     |s|={\rm arcosh}\left(\csc\left(\frac{\pi}{2+2z\nu r}\right)\right),
 \end{align}
that solely depends on the critical exponents $z\nu$ and the nonlinear exponent $r$. Moreover, since non-adiabatic excitations are formed in the vicinity of the critical point, only the nonlinear behavior of $g(t)$ close to $g_c$ is relevant~\footnote{Equivalent results can be obtained for different choices of $g(t)$ provided $|g_c-g(t)|\propto |\tau-t|^r$ for $|(\tau-t)/\tau|\ll 1$~\cite{sup}. See also~\cite{Barankov:08}.}. 


{\em Quantum critical battery.---} Let us consider a driven quantum harmonic oscillator as a quantum battery, which is loaded via the controllable creation of non-adiabatic excitations boosted at the quantum critical point. Through the cyclic transformation in Eq.~\eqref{eq:gt} we leave the battery in a state $\rho=|\psi(2\tau)\langle\psi(2\tau) \rangle |$ and stored a certain amount of work $\langle W\rangle$. 
Then, the maximum extractable work from the battery in a state $\rho$ under any cyclic unitary transformation is given by the so-called ergotropy~\cite{Allahverdyan:04,Francica:20}, given by $\mathcal{E}=\sum_n \epsilon_n(\rho_{nn}-r_n)$ where $H=\sum_n \epsilon_n \ket{n}\bra{n}$ with $\epsilon_n=n\omega$ in our case, and the state is $\rho=\sum_n r_n\ket{r_n}\bra{r_n}$ with $r_n$ the eigenvalues of $\rho$ sorted in descending order, and $\ket{r_n}$ the corresponding eigenstate, and $\rho_{nn}=\sum_{n'}r_{n'}|\langle r_{n'}|n\rangle|^2$.

\begin{figure}
\centering
\includegraphics[width=1\linewidth,angle=-0]{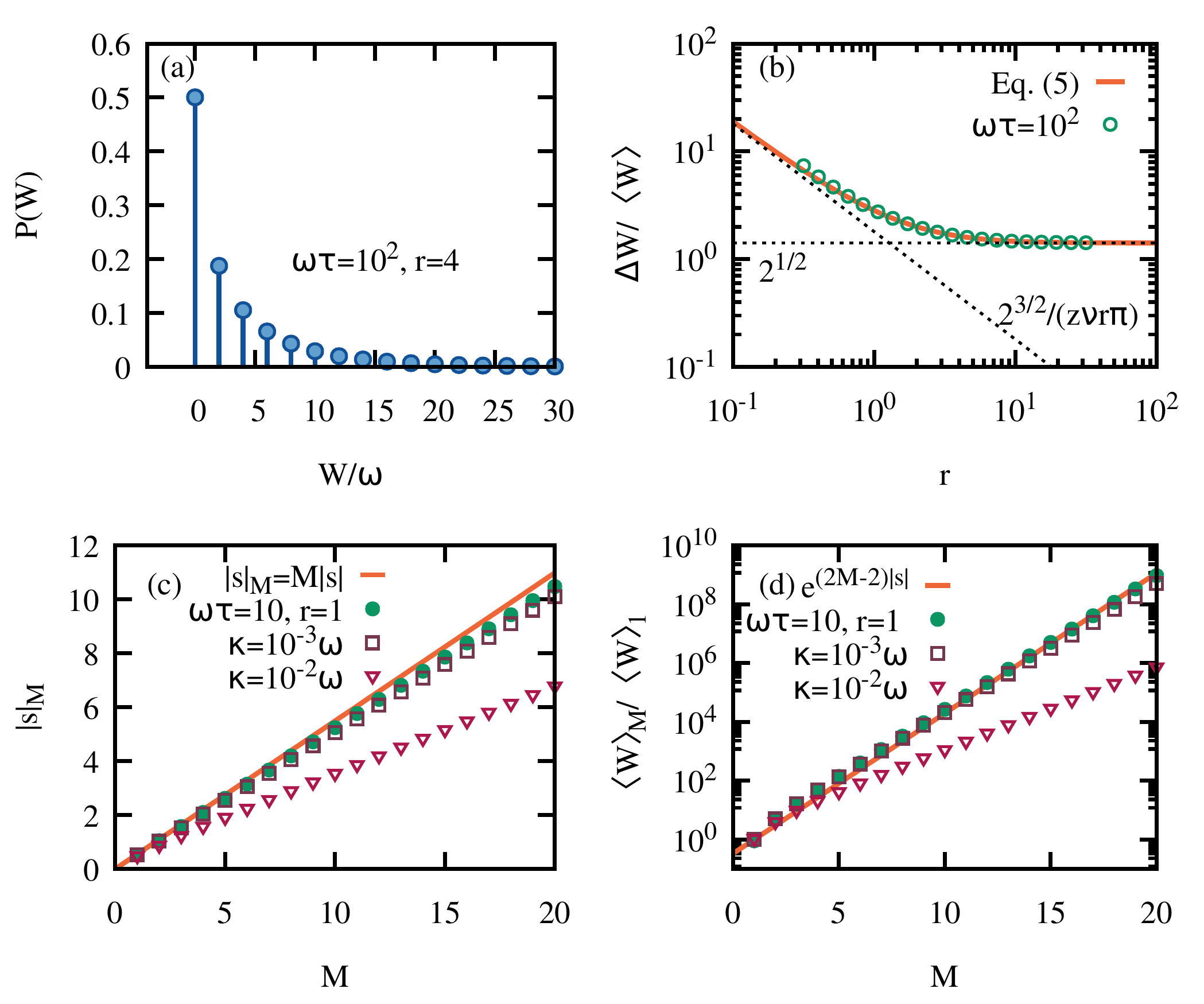}
\caption{\small{Quantum critical battery. Panel (a) shows an example for the work probability distribution $P(W)$ for $\omega\tau\!=\!10^2$ and $r\!=\!4$. The resulting fluctuations of the work distribution, $\Delta W/\langle W\rangle$, match the theoretical prediction (solid line), cf. Eq.~\eqref{eq:Aw}, as shown in panel (b) for $\omega\tau\!=\!10^2$. The dashed lines are guide to eyes for the asymptotic behavior, $\Delta W/\langle W\rangle\sim 2^{1/2}$ for $z\nu r\gg 1$, and $2^{3/2}/(z\nu r \pi)$ for $z \nu r\ll 1$. As illustrated in (c), repeating $M$ cycles can lead to an amplification of the squeezing parameter $|s|_M\!=\!M |s|$. This in turn implies an exponential increase of the stored and extractable work (cf. Eq.~\eqref{eq:WM}) $\langle W\rangle_M/\langle W\rangle_1\sim e^{(2M-2)|s|}$ as shown in (d). The square and diamond points in (c) and (d) show the decoherence effect on the battery. See main text for further details.}}
\label{fig2}
  \end{figure}

Upon completion of the protocol, the work performed on the battery follows a probability distribution given by $P(W)\!=\!\sum_{n=0} |c_n|^2\delta(W-n\omega)$, where $c_n=\langle n |\psi(2\tau)\rangle$. Since $\ket{\psi(2\tau)}\!=\!S(s)\ket{0}$ only even Fock states are populated, and thus energy is transferred to the battery in units of $2n\omega$ with $n=0,1,\ldots$. In addition, it is worth mentioning that $P(W)$ acquires a non-Gaussian profile. Since the amount of squeezing can be controlled by shaping the protocol $g(t)$, i.e. tuning the nonlinear exponent $r$, so are the average work and its variance, whose expressions are given by $\langle W\rangle\!=\!\omega\sinh^2(|s|)=\omega\tan^{-2}(\pi/(2+2z\nu r))$ and $\Delta^2 W=\langle W^2\rangle-\langle W\rangle^2=2\omega^2\cos^2(\pi/(2+2z\nu r))\sin^{-4}(\pi/(2+2z\nu r))$, respectively. Moreover, the ergotropy equals the stored work, $\mathcal{E}=\langle W\rangle$~\cite{sup}. In particular, the fluctuations of the work distribution, are given by
\begin{align}\label{eq:Aw}
    \frac{\Delta W}{\langle W\rangle}=\sqrt{2}\cos^{-1}\left(\pi/(2+2z\nu r)\right).
\end{align}
The fluctuations are dominant for $z\nu r\ll 1$, where $\Delta W/\langle W\rangle \sim 2^{3/2}(z\nu r\pi)^{-1}$, while they become constant in the opposite limit, $\Delta W/\langle W\rangle\sim 2^{1/2}$. The resulting work distribution and fluctuations are illustrated in Fig.~\ref{fig2}(a) and (b), respectively, which clearly shows how $\Delta W$ and $\langle W\rangle$, and thus $\mathcal{E}$, can be controlled by tailoring the nonlinear protocol. Note that the results plotted in Fig.~\ref{fig2}(a) and (b) are independent of the specific duration $\tau$ provided $\omega\tau\gtrsim 1$, i.e. excluding the regime of sudden quenches, which leave the state trivially unaltered.

So far we have focused on the realization of a single cycle. Yet, performing $M$
cycles can significantly boost the generated squeezing, and consequently the stored and extractable work beyond the linear scenario, $\langle W\rangle_M \gg M\langle W\rangle_1$ and $\mathcal{E}_M\gg M\mathcal{E}_1$. Indeed, the state after $M$ cycles under $g(t)$ can result in a $M$-fold squeezing, i.e. $|s|_M=M|s|$ with $|s|$ given in Eq.~\eqref{eq:s}. The total duration is then equal to $2M\tau$ and $g(2m\tau+t)\equiv g(t)$ with $m=0,1,\ldots M-1$. Remarkably, such proportional increase in $|s|_M$ translates into an exponential rise of the stored and extractable work~\cite{sup}, 
\begin{align}\label{eq:WM}
    \frac{\langle W\rangle_M}{\langle W\rangle_1}=\frac{\mathcal{E}_M}{\mathcal{E}_1}\sim e^{(2M-2)|s|},
\end{align}
 while the fluctuations enter the constant regime, $\Delta W_M/\langle W\rangle_M\sim 2^{1/2}$. This is reported in Fig.~\ref{fig2}(c) and (d). The solid points show the numerically-computed squeezing after the $M$th cycle, which follows well the predicted value $|s|_M= M|s|$. In this manner, the stored work $\langle W\rangle_M$ (and $\mathcal{E}_M$) grows exponentially with $M$, as exemplified in Fig.~\ref{fig2}(d) for $r=1$ and $\omega\tau=10$. Similar results can be found for other choices of $r$ and $\omega\tau$. 
 
 At this moment, a note is in order. Although for $M=1$ the amount of squeezing is $\tau$-independent (cf. Eq~\eqref{eq:s}), its phase naturally depends on the total evolution time. This accumulated phase becomes however relevant when $M>1$. As in any quantum engine or battery involving several cycles, the performance becomes phase sensitive~\cite{Watanabe:17,Reid:17}. In our case, we find that constructive interference of the accumulated phase leads to a sustained squeezing amplification across $M$ cycles when $\arg\{ b(2m\tau)\}=(2n+1)\pi/2$ with $n=0,1$ and $m=1,2,\ldots,M$~\cite{sup}. To the contrary, if $\arg\{b(2\tau)\}=2n\pi$ with $n=0,1$, the subsequent cycle ($M=2$) counteracts leading to a suppression of squeezing so that $|s|_{M=2}\approx 0$, and $|s|_M\lesssim |s|$. In particular, we find $\arg\{b(2\tau)\}={\rm mod}\{-\pi(1+\omega\tau)/2,2\pi\}$ for $r=1$ so that $\omega\tau=10$ leads to $\pi/2$ and holds for increasing $M$. This case corresponds to Fig.~\ref{fig2}(c) and (d) (see~\cite{sup} for further details). Finally, we comment that decoherence impairs the performance, although an exponential advantage can still be achieved if $2\tau \kappa\ll 1$ where $\kappa$ denotes the noise rate (cf. Fig.~\ref{fig2}(c) and (d)).  For that, we consider a $T=0$ reservoir interacting with the system in a standard Lindblad form~\cite{Breuer}, $\dot{\rho}=-i[H(t),\rho]+\kappa/2(2a\rho a^\dagger-\{a^\dagger a ,\rho\})$ (see~\cite{sup} for details).


{\em Spin squeezing and metrological gain.---} Let us now consider N spin-$1/2$ particles collectively coupled and interacting as described by the LMG Hamiltonian~\cite{Lipkin:65}, which can be written as
\begin{align}\label{eq:LMG}
    H_{\rm LMG}(t)=-\omega J_z-\frac{g^2(t)\omega}{N}J_x^2,
\end{align}
where the pseudo-spin operators are defined as $J_\alpha=\frac{1}{2}\sum_{i=1}^N\sigma^\alpha_i$ for $\alpha\in\{x,y,z\}$, where $\sigma^\alpha_i$ refer to the Pauli matrices for the $i$th spin. Since $[J^2,H_{\rm LMG}]=0$ one can restrict to the highest pseudo-spin subspace, where the dimension of the Hilbert space grows linearly with $N$. 

In the thermodynamic limit, $N\rightarrow\infty$, the LMG exhibits a QPT at $g_c=1$~\cite{Ribeiro:07,Ribeiro:08}. In this limit, the LMG can be effectively described by a single bosonic mode. This is achieved by performing the Holstein-Primakoff transformation, $J_z=J-a^\dagger a$ and $J_+=\sqrt{2J}\sqrt{1-a^\dagger a/(2J)}a$ and $J_x=(J_++J_-)/2$. Taking the $N\rightarrow\infty$ limit, $H_{\rm LMG}$ reduces to Eq.~\eqref{eq:H0} up to a constant energy shift, and therefore $z\nu=1/2$~\cite{Ribeiro:07,Ribeiro:08}.
Thus, the bosonic squeezed state $S(s)\ket{0}$ translates into a spin squeezing of the form
\begin{align}\label{eq:sspin}
    \ket{\xi}=S_{\rm spin}(\xi)\ket{J,m_J=J}
\end{align}
where $S_{\rm spin}(\xi)=e^{\xi^*J_+^2/2-\xi J_-^2/2}$ corresponds to the spin squeezing operator. From the previous considerations, one can immediately see that $|\xi|=|s|/N$ in the thermodynamic limit, with $|s|$ given in Eq.~\eqref{eq:s}. On the one hand, as the system size increases $N\rightarrow\infty$, i.e. as finite-size effects become negligible, the evolved state upon a cycle $g(t)$, $\ket{\psi(2\tau)}=U(2\tau)\ket{J,m_J=J}$ with $U(t)=\mathcal{T}e^{-i\int_0^t ds H_{\rm LMG}(s)}$, becomes closer to $\ket{\xi}$ with $|\xi|=|s|/N$. 

On the other hand, since the energy gap $\epsilon(g_c)$ of $H_{\rm LMG}$ is non-zero for any finite $N$, the duration of the protocol $\tau$ must be such that the dynamics lie in the quasiadiabatic regime. That is, $\omega\tau\gtrsim 1$ as aforementioned (to exclude sudden quench dynamics), together with $\epsilon(g_c)\tau \lesssim 1$, which prevents true adiabatic cycles in a finite system. Since $\epsilon(g_c)\propto N^{-z}$ with $z=1/3$~\cite{Dusuel:04,Dusuel:05}, it follows that $1\lesssim\omega\tau\lesssim N^{1/3}$. 
In addition, it is worth mentioning that for half a cycle, i.e. for a ramp toward the critical point, Kibble-Zurek scaling laws emerge in this driving regime~\cite{Hwang:15,Puebla:17,Defenu:18,Puebla:20}, which further highlights that non-adiabatic excitations are caused due to the existence of a critical point.

\begin{figure}
\centering
\includegraphics[width=1\linewidth,angle=-0]{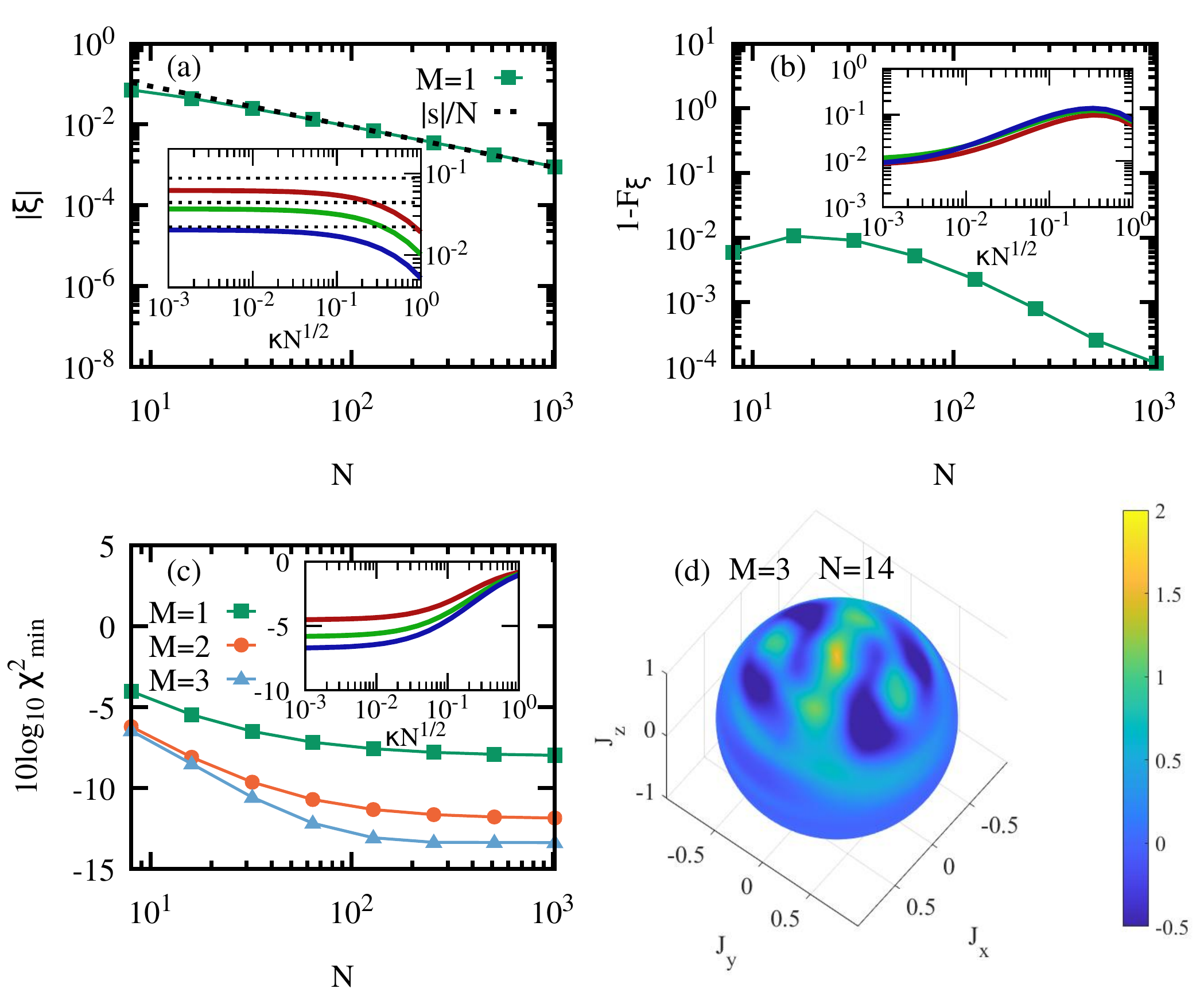}
\caption{\small{Spin squeezing and metrological gain. Panel (a) shows the resulting squeezing parameter $|\xi|$ for $\kappa=0$ as a function of the number of spins $N$ for one cycle. The dashed line shows the predicted theoretical value $|s|/N={\rm arcosh}\left(\csc(\pi/(2+2z\nu r))\right)/N$ with $z\nu=1/2$, while the inset shows $|\xi|$ as $\kappa$ increases for $N=10$, $20$ and $40$ spins (lines from top to bottom) with their corresponding $|s|/N$ (dashed lines). The fidelity of the generated state is very close to the expected spin squeezed state, as shown in panel (b) for $\kappa=0$, i.e. $F_\xi\rightarrow 1$ as $N\rightarrow\infty$. The inset shows $F_{\xi}$ for $\kappa\neq 0$ as in (a). In (c) we show $\chi^2_{\rm min}$ (cf. Eq.~\eqref{eq:chi}) for $M=1$, $2$ and $3$ cycles which demonstrates the amount of useful multipartite entanglement, and survives for $\kappa\neq 0$ (see inset for $M=1$ cycle). Panel (d) shows the Wigner function $\mathcal{W}(\theta,\phi)$ for the state achieved upon $M=3$ cycles, $N=14$ spins and $\kappa=0$. All the results correspond to $\omega\tau=2$ and $r=2$. See main text for further details.}}
\label{fig3}
  \end{figure}
In order to exemplify the generation of spin squeezed states via a cyclic protocol $g(t)$, we compute the state of the form in Eq.~\eqref{eq:sspin} which maximizes $F_\xi=\langle \xi|\rho|\xi\rangle$ for different system sizes $N$ and where $\rho$ denotes the state after the cycle according to $\dot{\rho}=-i[H_{\rm LMG}(t),\rho]+\kappa/2 (J_+\rho J_--\{J_+J_-,\rho\})$. For $\kappa=0$, the resulting parameter $|\xi|$ closely follows the expected relation $|\xi|=|s|/N$, as illustrated in Fig.~\ref{fig3}(a) for $\omega\tau=2$ and $r=2$, which also holds for reasonably small noise rates $\tau\kappa\sqrt{N}\lesssim 10^{-2}$. Moreover, the fidelity $F_\xi\rightarrow 1$ as $N\rightarrow\infty$, which is shown in Fig.~\ref{fig3}(b). For $N=10^3$ we find already $F_\xi\approx 0.9999$, which corroborates the ability to generate spin squeezing by harnessing controllable non-adiabatic excitations generated by a QPT. Note that slower cycles impair the generation of spin squeezing~\cite{sup}.

Such spin squeezed states may contain multipartite entanglement shared by its $N$ spins and thus, they may offer a metrological advantage for sub shot-noise phase sensitivity. In order to quantify such advantage, we rely on the quantity $\chi^2_{\rm min}$ related to the quantum Fisher information~\cite{Pezze:09}, which witnesses multipartite entanglement and metrological advantage (sub shot-noise sensitivity) when $\chi^2_{\rm min}<1$, and reads 
\begin{align}\label{eq:chi}
    \chi^2_{\rm min}=\min_{{\vec{n}}}\frac{N}{4 (\Delta R_{\vec{n}})^2},
\end{align}
with the operator $R_{\vec{n}}$ defined as $\{R_{\vec{n}},\rho\}=i[J_{\vec{n}},\rho]$, $(\Delta R_{\vec{n}})^2=\langle R_{{\vec n}}\rangle^2-\langle R_{\vec{n}}\rangle^2$ its variance on the state $\rho$ and  $J_{\vec{n}}=(J_x\sin\theta\cos\phi,J_y\sin\theta\sin\phi,J_z\cos\theta)$ being the pseudo-spin operator in an arbitrary direction ${\vec{n}}$. Note that $\Delta R_{\vec{n}}=\Delta J_{\vec{n}}$ for a pure state $\rho$. From Eq.~\eqref{eq:sspin} it is clear that $\chi^2_{\rm min}$ is achieved when $\theta=\pi/2$, while the angle $\phi$ in the $xy$-plane depends on the phase of $\xi$. We compute $\chi^2_{\rm min}$ minimizing over any possible pair of angles $\{\theta,\phi\}$ for each state in Fig.~\ref{fig3}(a) and find that the produced states always contain multipartite entanglement, provided $\tau\kappa \sqrt{N}\lesssim 1$, and are thus useful for sub shot-noise sensitivity since $\chi^2_{\rm min}<1$.  Moreover, as shown above in the bosonic case, realizing more cycles may lead to more squeezing. In this context, this is reflected in a reduction of $\chi^2_{\rm min}$, as shown in Fig.~\ref{fig3}(c), where $\chi^2_{\rm min}\approx 0.05$ after $M=3$ cycles for $N=10^3$ spins. However, further cycles do not improve significantly $\chi^2_{\rm min}$. In order to illustrate the features of the resulting spin state, we show as an example the Wigner function $\mathcal{W}(\theta,\phi)$ for $N=14$ spins upon $M=3$ cycles, which is calculated following the standard procedure~\cite{Dowling:94,sup}, while the examples for $M=1$ and $2$ are provided in~\cite{sup}. 

{\em Conclusions.---} In this Letter, we have investigated the production of non-adiabatic excitations as a result of cyclic protocols reaching a quantum critical point, showing how they can be controlled and thus harnessed. In particular, we have focused on critical fully-connected models, whose effective description reduces to a driven quantum harmonic oscillator. As a result of the breakdown of the adiabatic condition, the amount of non-adiabatic excitations depends solely on the critical exponents and the shape of the protocol close to the QPT. These non-adiabatic excitations produce squeezing in the initially-prepared ground state. 

We first considered an effective model, namely, a quantum harmonic oscillator as a quantum battery, where work is performed through the cyclic driving, whose average value and fluctuations obey simple relations and are dictated by the nonlinear form of the protocol in the vicinity of the critical point, as well as the maximum extractable work. Interestingly, we showed that the realization of $M$ consecutive cycles can yield an exponential increase of the stored and extractable work. 

As a second example, we considered a system made of $N$ spin-$1/2$ particles interacting according to the LMG model, which reduces to an effective quantum harmonic oscillator model in the thermodynamic limit. We showed that non-adiabatic excitations translate into spin squeezing. In this manner, by performing a quasiadiabatic cycle toward the critical point, one generates spin squeezing in a controllable fashion. Moreover, such spin squeezed states contain multipartite entanglement and are therefore useful for sub shot-noise phase sensitivity. 

Finally, since different and experimentally realizable fully-connected quantum many-body systems can be effectively described in terms of a quantum harmonic oscillator~\cite{Anquez:16,Mottl:12,Brennecke:13,Zibold:10,Jurcevic:17,Cai:21}, our findings are relevant to distinct platforms. Our results will motivate further research in the exciting arena of nonequilibrium phenomena and critical dynamics with applications in quantum thermodynamics and quantum metrology.


\begin{acknowledgements}
 This work has been supported by the European Union's Horizon 2020 FET-Open project SuperQuLAN (899354), H2020-FET-OPen 2018-2020 TEQ (766900), Royal Society Wolfson Research Fellowship (RSWF/R3/183013),  Leverhulme Trust Research Project Grant (RGP-2018-266), SFI-DfE Investigator Programme (15/IA/2864), UK EPSRC EP/S02994X/1, UK EPSRC EP/T026715/1 and UK EPSRC EP/T028106/1. 
\end{acknowledgements}


%

\widetext
\clearpage
\begin{center}
\textbf{\large Supplemental Material \\ Harnessing non-adiabatic excitations promoted by a quantum critical point}
\end{center}
\setcounter{equation}{0}
\setcounter{figure}{0}
\setcounter{table}{0}
\setcounter{page}{1}
\makeatletter
\renewcommand{\theequation}{S\arabic{equation}}
\renewcommand{\thefigure}{S\arabic{figure}}
\renewcommand{\bibnumfmt}[1]{[S#1]}
\renewcommand{\citenumfont}[1]{S#1}

\begin{center}
  Obinna Abah${}^1$, Gabriele De Chiara${}^{1}$, Mauro Paternostro${}^1$, and Ricardo Puebla${}^{2,1}$\\
    \vspace{0.2cm}{\small ${}^1${\em Centre for Theoretical Atomic, Molecular, and Optical Physics,\\ School of Mathematics and Physics, Queen's University, Belfast BT7 1NN, United Kingdom}}\\
    \vspace{0.0cm}{\small $^2${\em Instituto de F{\'i}sica Fundamental, IFF-CSIC, Calle Serrano 113b, 28006 Madrid, Spain}}
  \end{center}

\section*{I. Details for the dynamics of the quantum critical harmonic oscillator}
Here we provide more details of the derivation for the critical dynamics summarized in the main text. Let us consider the driven quantum harmonic oscillator,
\begin{align}\label{eqSM:Ht}
    H(t)=\omega a^\dagger a-\frac{g^2(t)\omega}{4}(a+a^\dagger)^2,
\end{align}
following a general nonlinear cycle $g(t)$. We can exploit the quadratic nature of the Hamiltonian and solve the Schr{\"o}dinger equation for the evolution operator, $\dot{U}(t)=-iH(t)U(t)$. In particular, neglecting an irrelevant global phase, the equation of motion for $U(t)$ reads as
\begin{align}\label{eq:Ut}
    \dot{U}(t)=-i\omega \left( \left(1-\frac{g^2(t)}{2}\right)a^\dagger a-\frac{g^2(t)}{4}(a^2+(a^{\dagger})^2)\right)U(t),
\end{align}
Using the coherent state representation, $\ket{\alpha}=D(\alpha)\ket{0}$ where $D(\alpha)$ is the displacement operator, we introduce $U_\alpha(t)=\bra{\alpha}U(t)\ket{\alpha}$ so that
\begin{align}
    \dot{U}_{\alpha}(t)=-i\omega \left( \left(1-\frac{g^2(t)}{2}\right)\alpha^*(\alpha+\partial_{\alpha^*})-\frac{g^2(t)}{4}\left(\left(\alpha+\partial_{\alpha^*}\right)^2+\alpha^{*,2}\right) \right)U_{\alpha}(t),
\end{align}
which suggests a solution of the form of 
\begin{align}
    U_\alpha(t)={\rm exp}\left[k(t)+b(t)\alpha^{*,2}+c(t)\alpha^*\alpha+d(t)\alpha^2\right].
\end{align}
The equations of motion for the coefficients $k(t)$, $b(t)$, $c(t)$ and $d(t)$ follow from Eq.~\eqref{eq:Ut}. In particular, $b(t)$ is decoupled from the rest
\begin{align}\label{eqSM:bt}
    \dot{b}(t)=-i\omega\left(-\frac{g^2(t)}{4}+2\left(1-\frac{g^2(t)}{2}\right)b(t)-g^2(t)b^2(t)\right),
\end{align}
which can be recast as Eq. (3) in the main text. 
Hence, we can express the operator $U_\alpha(t)$ in its original form by replacing $\alpha$ and $\alpha^*$ by $a$ and $a^\dagger$ requiring normal ordering,
\begin{align}
    U(t)=e^{k(t)}e^{b(t)(a^{\dagger})^2}N[e^{c(t)a^\dagger a}]e^{d(t)a^2}.
\end{align}
Note that $e^{k(t)}$ is just a constant prefactor that ensures that the state is normalized. In particular, $k(t)$ obeys $\dot{k}(t)=i\omega g^2(t) b(t)/2$. In this manner, an evolved state under the protocol $g(t)$ is given by
\begin{align}
    \ket{\psi(t)}=U(t)\ket{\psi(0)},
\end{align}
where we have assumed $t=0$ as initial time. It is easy to see now that if the initial state is the vacuum, $\ket{\psi(0)}=\ket{0}$, then 
\begin{align}
    \ket{\psi(t)}=e^{k(t)}e^{b(t)(a^{\dagger})^2}\ket{0} \propto e^{b(t)(a^{\dagger})^2}\ket{0},
\end{align}
up to an irrelevant global phase and with initial condition $b(0)=0$ in Eq.~\eqref{eqSM:bt}. 

We further note that the squeezing operator $S(s)=e^{s a^{\dagger,2}/2-s^* a^2/2}$ can be split in a standard way as~\cite{Fisher:84SM,Truax:85SM}
\begin{align}
    S(s)={\rm exp}\left[\frac{1}{2}(e^{i\theta}\tanh|s|)(a^{\dagger})^2\right] {\rm exp}\left[-\log(\cosh|s|)(a^\dagger a+\frac{1}{2})\right] {\rm exp}\left[-\frac{1}{2}(e^{i\theta}\tanh|s|)a^2\right], 
\end{align}
which immediately leads to the identification
\begin{align}\label{eq:sbt}
    |s|&={\rm artanh}(2|b(t)|)\\
    \theta&=\arg\{{b(t)}\}.
\end{align}
Clearly, it also follows that $|e^{k(t)}|=(\cosh(|s|)^{-1/2}$ with $|s|$ as given in Eq.~\eqref{eq:sbt}. 

On the other hand, the dynamics of a driven quantum harmonic oscillator (Eq.~\eqref{eqSM:Ht} or Eq. (1) in the main text) can be solved using the so-called Ermakov equation~\cite{Lewis:67SM,Lewis:68SM,Lewis:69SM},
\begin{align}
    \ddot{\xi}(t)=\frac{1}{4\xi^3(t)}-\omega^2(t)\xi(t),
\end{align}
 with $\xi(0)=(2\omega(0))^{-1/2}$ and $\omega^2(t)=\omega^2(1-g^2(t))$. Following the recent development by N. Defenu~\cite{Defenu:20SM}, one can find an exact solution for this problem in the long-time limit which unveils the breakdown of the adiabatic condition. In particular, the overlap of the evolved state upon a cycle with a nonlinear protocol $g(t)$ is given by
 \begin{align}\label{eq:fznur}
     \lim_{\tau\rightarrow \infty}|\langle 0|\psi(2\tau)\rangle|^2=\sin\left( \frac{\pi}{2+2z\nu r}\right),
 \end{align}
which holds approximately for finite $\tau$ provided $\omega\tau\gtrsim 1$ (i.e. so that excitations are mainly caused in the vicinity of the critical point and excluding the sudden quench limit, $\tau\rightarrow 0$, which leaves trivially unaltered the initial state). 

\begin{figure}
\centering
\includegraphics[width=0.9\linewidth,angle=-0]{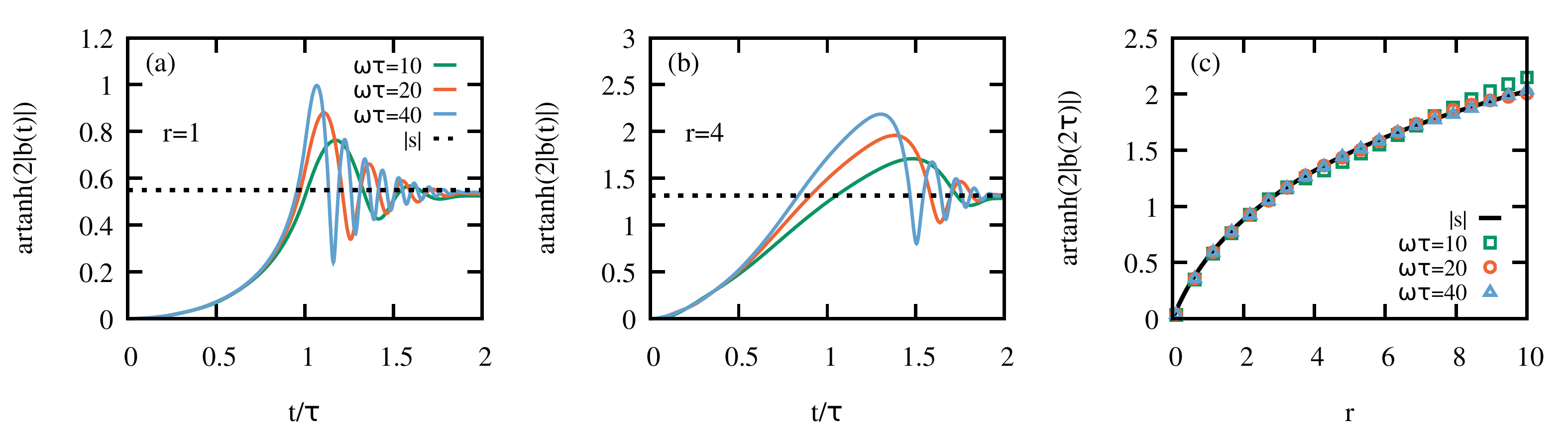}
\caption{\small{Panels (a) and (b) show the dynamics of the parameter $b(t)$ for different protocols, $r=1$ (a) and $r=4$ (b), and for three different duration $\omega\tau=10$, $20$ and $40$. The dashed line corresponds to the predicted value $|s|={\rm arcosh}(\csc(\pi/(2+2z\nu r)))$. The actual squeezing ${\rm artanh}(2|b(2\tau)|)$ at the end of the cycle converges to the predicted value. Panel (c) illustrates the dependence of $|s|$ on the nonlinear exponent $r$ (solid line), together with the numerically computed squeezing for protocols with different duration.}}
\label{figb}
\end{figure}

Note that $|\bra{0}S(s)\ket{0}|^2=\cosh^{-1}(|s|)$, and thus combining Eqs.~\eqref{eq:sbt} and~\eqref{eq:fznur}, we finally obtain
\begin{align}\label{eq:stheo}
    |s|={\rm arcosh}\left(\csc\left(\frac{\pi}{2+2z\nu r}\right)\right),
\end{align}
or equivalently,
\begin{align}
    |b(2\tau)|\approx\frac{1}{2}\cos\left(\frac{\pi}{2+2z\nu r}\right),
\end{align}
where the approximate character of the previous expression refers to small deviations due to a finite time $\tau$. See Fig.~\ref{figb} for examples of the squeezing parameter ${\rm artanh}(2|b(t)|)$ for different cycle durations $\tau$ and distinct nonlinear protocols, $r=1$ and $r=4$ in Fig.~\ref{figb}(a) and (b), respectively. The numerically computed evolutions ${\rm artanh}(2|b(t)|)$ match the predicted value $|s|={\rm arcosh}(\csc(\pi/(2+2z\nu r)))$ to a very good approximation (cf. Eq.~\eqref{eq:stheo}). This is further corroborated in Fig.~\ref{figb}(c) where the final squeezing ${\rm artanh}(2|b(2\tau)|)$ is plotted as a function of the nonlinear exponent $r$, revealing a very good agreement with the theoretical prediction.

\begin{figure}
\centering
\includegraphics[width=0.6\linewidth,angle=-0]{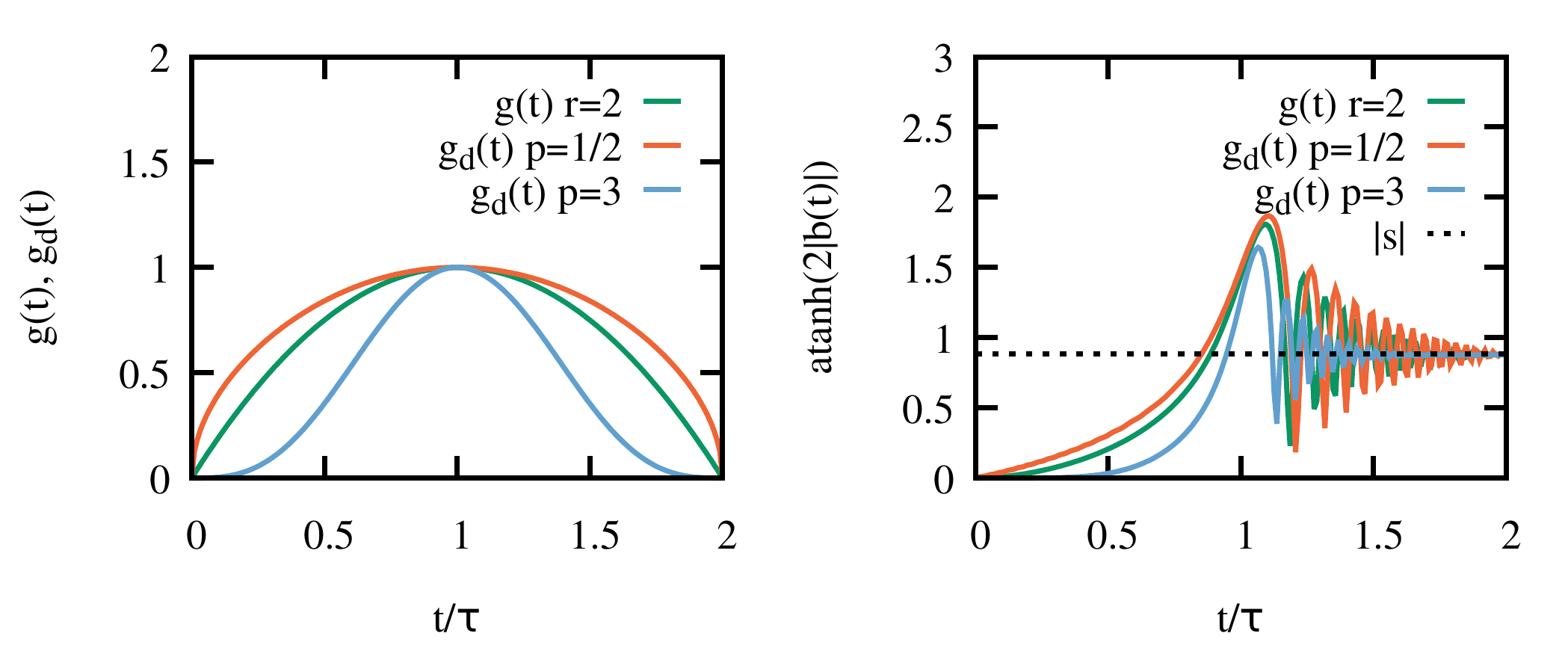}
\caption{\small{Panel (a) shows the protocol $g(t)$ with $r=2$, used in the main text, and two examples of the different protocol $g_d(t)$, as given in Eq.~\eqref{eq:gdt}, with $p=1/2$ and $3$, while panel (b) corresponds to the actual dynamics of the squeezing parameter ${\rm artanh}(2|b(t)|)$ under these distinct protocols for $\omega\tau=10^2$. The dashed line corresponds to $|s|={\rm arcosh}(\csc(\pi/(2+2z\nu r)))\approx 0.8814$ since $r=2$ and $z\nu=1/2$. Note that all converge to the same value after the cycle is completed.}}
\label{figdifprot}
\end{figure}

\section*{II. Analysis of different protocols}
As stated in the main text, the results are robust with respect to the specific shape of the protocol provided $g(t)$ behaves in a nonlinear fashion in the vicinity of the critical point $g_c$. That is, the actual shape of $g(t)$ away from $g_c$ is irrelevant and only the functional form close to $g_c$ dictates the resulting squeezing $|s|$. In order to illustrate this, we consider a different protocol $g_d(t)$
\begin{align}\label{eq:gdt}
    g_d(t)=\begin{cases} &g_c\sin^p\left(\frac{\pi}{2}\frac{t}{\tau} \right)\quad 0\leq t\leq \tau \\
    & g_c\cos^p\left(\frac{\pi}{2}\frac{t-\tau}{\tau}\right)\quad \tau<t\leq 2\tau,
    \end{cases}
\end{align}
with $p>0$. One can immediately verify that 
\begin{align}
    g_c-g_d(t)= \frac{1}{8}(\pi^2 p)\frac{|t-\tau|^2}{\tau^2}+O\left(\frac{|t-\tau|^4}{\tau^4}\right).
\end{align}
In this manner, $g_d(t)$ behaves as $g(t)$ with $r=2$ close to $g_c$ since $g_c-g(t)\approx |t-\tau|^2$ for $|t-\tau|\ll 1$.
 Hence, although the actual dynamics following $g_d(t)$ does depend on $p$, and will be different with respect to the one under $g(t)$, the resulting non-adiabatic excitations, promoted in the vicinity of $g_c$, are equivalent to those obtained following $g(t)$ with $r=2$, regardless of the specific value of $p$. Note that this is similar to the resulting Kibble-Zurek scaling laws when nonlinear protocols are considered~\cite{Barankov:08SM}. This means that $|b(2\tau)|$ matches that of $g(t)$ with $r=2$, independently of $p$, namely, ${\rm artanh}(2|b(2\tau)|)=|s|={\rm arcosh}(\csc(\pi/(2+2z\nu r)))\approx 0.8814$ since $z\nu=1/2$. This example is plotted in Fig.~\ref{figdifprot}, but we stress that the same behavior holds for other protocols.

\section*{III. Repeating cycles}
Here we provide more details regarding the realization of $M$ consecutive cycles under a nonlinear protocol reaching the quantum critical point. We first note that, while the primary cycle leads to a robust squeezing as detailed in the previous section, subsequent cycles will be susceptible to the accumulated phase during previous cycles. That is, the initial condition for the $M+1$ cycle is given by $b(2M\tau)$ resulting from completing the $M$ cycle. Thus, depending on the complex phase, $\theta\equiv \arg\{b(2M\tau)\}$, the cycle may lead to an amplification of the total squeezing, i.e., to an increasing $|b(2(M+1)\tau)|$ value. Due to the complex and nonlinear nature of Eq.~\eqref{eqSM:bt}, we rely on numerical simulations, and find that a sustained amplification can be achieved whenever $\theta=(2n+1)\pi/2$ with $n=0,1$, so that $|s|_M\approx M|s|$. On the other hand, destructive interference occurs for $\theta=2n\pi$ with $n=0,1$. This accumulated phase during the cycle can be expressed as $\theta={\rm mod}\{-\Theta(\tau)-\pi/2,2\pi\}$ where $\Theta(\tau)=\int_0^{2\tau}dt \epsilon(g(t))$ and $\epsilon(g)=\omega\sqrt{1-g^2(t)}$. For $r=1$, one finds that $\Theta(\tau)=\pi\omega\tau/2$ which leads to the expression given in the main text. In this manner, $\omega\tau=10$ and $r=1$ gives $\theta=\pi/2$, while $\omega\tau=11$ brings the phase to $\theta=0$. 
Note that $\Theta(\tau)$ (and thus $\theta$) depends non-trivially on $r$, e.g., for $r=2$ it results in $\Theta(\tau)=2(2\sqrt{2}-1)\omega\tau/3$.

\begin{figure}
\centering
\includegraphics[width=0.9\linewidth,angle=-0]{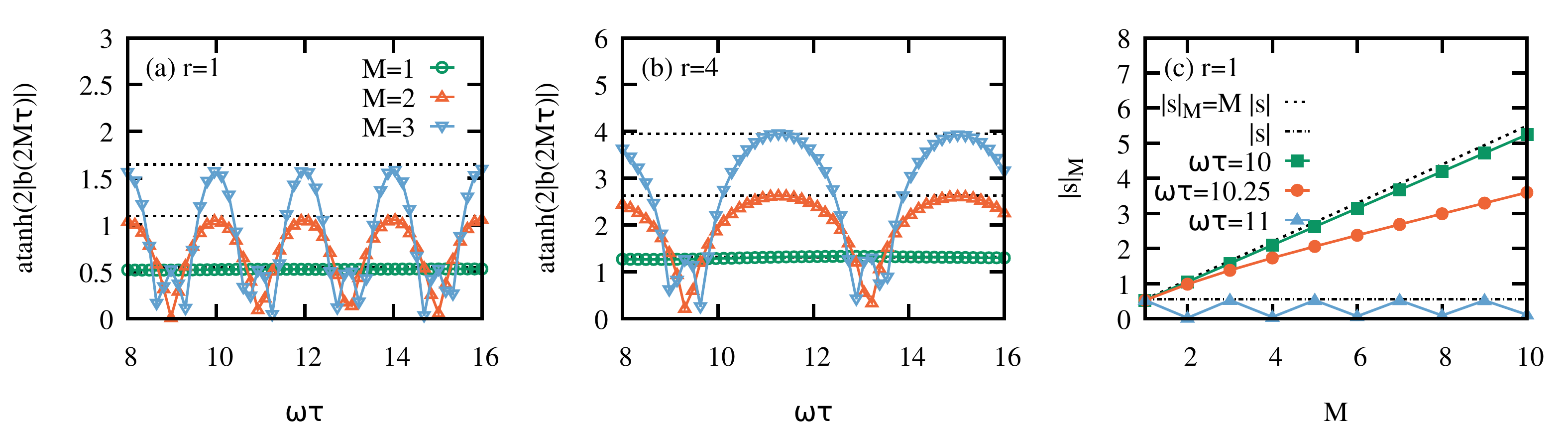}
\caption{\small{Squeezing parameter after $M$ cycles. The panels show the squeezing after the cycle $M=1$, $2$ and $3$, as a function of the protocol duration $\tau$, which is given by ${\rm artanh}(2|b(2M\tau)|)$, for $r=1$ (a) and $r=4$ (b). While for $M=1$ the squeezing does not depend on $\tau$, the phase gained during the cycle is relevant for further amplification, which is reflected in the oscillations for $M=2$ and $M=3$. For certain $\omega\tau$ the amplification can be sustained for increasing $M$ (e.g. $\omega\tau=10$ for $r=1$, which corresponds to the Fig. 2(c) and (d) of the main text; see also panel (c)). Dashed lines correspond to the predicted amplified squeezing, $|s|_M=M|s|=M{\rm arcosh}(\csc(\pi/(2+2z\nu r)))$.  Panel (c) shows the resulting squeezing $|s|_M$ for different choices of $\omega\tau$ with $r=1$, and as a function of the number of cycles $M$. For $\omega\tau=10$ we obtain $|s|_M\approx M |s|$ (cf. Fig. 2(c) in main text), while for $\omega\tau=11$ it results in a destructive interference, which leads to $|s|_{2M+1}\approx |s|$ (odd cycles) and $|s|_{2M}\approx 0$ (even cycles).}}
\label{figcycles}
\end{figure}

In Fig.~\ref{figcycles} we show the typical profile for the squeezing application of consecutive $M$ cycles. While the primary cycle $M=1$ does not depend on the duration $\tau$, the phase $\theta$ is crucial to accomplish a sustained amplification of the squeezing parameter for $M>1$ due to interference effects.  As aforementioned, one finds $\theta=\pi/2$ for $\omega\tau=10$ and $r=1$. This case corresponds to the results plotted in Fig.~2(c) and (d) of the main text, which clearly demonstrate that $|s|_M=M|s|$.  To the contrary, if $\omega\tau$ is such that $\theta=2n\pi$ after the first cycle, the following one counteracts the generated squeezing so that $|s|_{M=2}\approx 0$, as it can be seen in Fig.~\ref{figcycles} (for $\omega\tau=11$). In the latter case, no amplification is achieved and the repetition of $M$ cycles is only capable to reach a maximum squeezing given by $|s|$ (cf. Fig.~\ref{figcycles}). 

For $|s|_M=M|s|$, we have
\begin{align}
    \langle W\rangle_M=\omega\sinh^2(|s|_M)=\omega\sinh^2(M|s|),
\end{align}
and thus it is easy to see that
\begin{align}\label{eqSM:WM}
    \frac{\langle W\rangle_M}{\langle W\rangle_1}=\frac{e^{2M|s|}+e^{-2M|s|}-2}{e^{2|s|}+e^{-2|s|}-2}\sim e^{(2M-2)|s|}.
\end{align}
On the other hand, since the state is a squeezed vacuum state, $|\psi(t)\rangle=S(s(t))\ket{0} $, the expression fo the ergotropy can be easily computed and equals the stored energy $\mathcal{E}=\langle W\rangle$. Recall that $\mathcal{E}=\sum_n \epsilon_n (\rho_{nn}-|c_n|^2)$ where $H=\sum_n \epsilon_n \ket{n}\bra{n}$ and $\epsilon_n<\epsilon_{n+1}$, $\ket{n}$ the $n$th Fock state, and $\rho=\sum_n r_n \ket{r_n}\bra{r_n}$ with $r_n\geq r_{n+1}$, and $\rho_{nn}=\sum_{n'}r_{n'} |\langle r_{n'}|n\rangle|^2$. In our case $\epsilon_n=n\omega$, while $r_0=1$ ($r_{n>0}=0$), $\ket{r_0}=S(s)\ket{0}$, and therefore $|\langle r_{0}|2n\rangle|^2=\cosh^{-1}(|s|)(\tanh^2(|s|)(2n!)/(2^n \ n!)^2$, together with $|\langle r_0|2n+1\rangle|^2=0$. In this manner, the ergotropy takes the form $\mathcal{E}=\sum_n \epsilon_n(\rho_{nn}-r_n)=\langle W\rangle -\sum_n n\omega r_n =\langle W\rangle$. 
Hence, $\mathcal{E}_M=\langle W\rangle_M$, so that $\mathcal{E}_M/\mathcal{E}_1\sim e^{(2M-2)|s|}$,
which together with~\eqref{eqSM:WM} corresponds to the expression in Eq. (6) of main text.

\section*{IV. Noisy critical cycles}
In order to simulate the master equation $\dot{\rho}=-i[H(t),\rho]+\kappa/2(2a\rho a^\dagger-\{a^\dagger a,\rho\})$ we rely on the Wigner characteristic function $\chi(\beta,\beta^*,t)={\rm Tr}[e^{\beta a^\dagger-\beta^* a}\rho(t)]$ with $\beta,\beta^* \in \mathbb{C}$. Exploiting the quadratic nature of the master equation, the Fokker-Planck equation of motion $\dot{\chi}(\beta,\beta^*,t)$ admits a Gaussian ansatz, $\chi(\beta,\beta^*,t)=e^{i{\bf u}^t \mu(t)-(1/2){\bf u}^t \Sigma(t){\bf u}}$ with ${\bf u}^t=(\beta,\beta^*)$, and $\mu(t)$ and $\Sigma(t)$ the first and second moments, respectively. Since $\mu(0)=0$ and the master equation contains only quadratic terms, it follows $\mu(t)=0 \ \forall t$. Defining $2\sigma(t)={\rm Tr}[\Sigma(t)]=\sigma_{00}(t)+\sigma_{11}(t)$, the master equation reduces to
\begin{align}\label{eqSM:Gauss}
    \dot{\sigma}(t)&=\kappa(1/2-\sigma(t))+ig^2(t)\omega (\sigma_{01}(t)-\sigma_{10}(t))/2\\
    \dot{\sigma}(t)&=(2i\omega-ig^2(t)\omega-\kappa)\sigma_{10}(t)+ig^2(t)\omega\sigma(t)
\end{align}
where $\sigma_{01}(t)=\sigma_{10}^*(t)$ are the non-diagonal elements of $\Sigma(t)$. The initial condition is ${\rm Tr}[a^\dagger a \rho(0)]=\sigma(0)-1/2=0$ and $\sigma_{01}(0)=0$. From the quadratures of $a$ and $a^\dagger$, we can identify the squeezing at the end of the evolution as $|s|={\rm arsinh}(|2\sigma_{01}(2\tau)|)/2$, while the performed work $\langle W\rangle={\rm Tr}[H(2\tau)\rho(2\tau)]=\omega{\rm Tr}[a^\dagger a \rho(2\tau)]=\omega(\sigma(2\tau)-1/2)$.  The numerical results obtained solving Eq.~\eqref{eqSM:Gauss} for different values of $\kappa$ are plotted in Fig.~2 of the main text. 

\section*{V. Spin squeezing in the LMG model}
\begin{figure}
\centering
\includegraphics[width=0.6\linewidth,angle=-0]{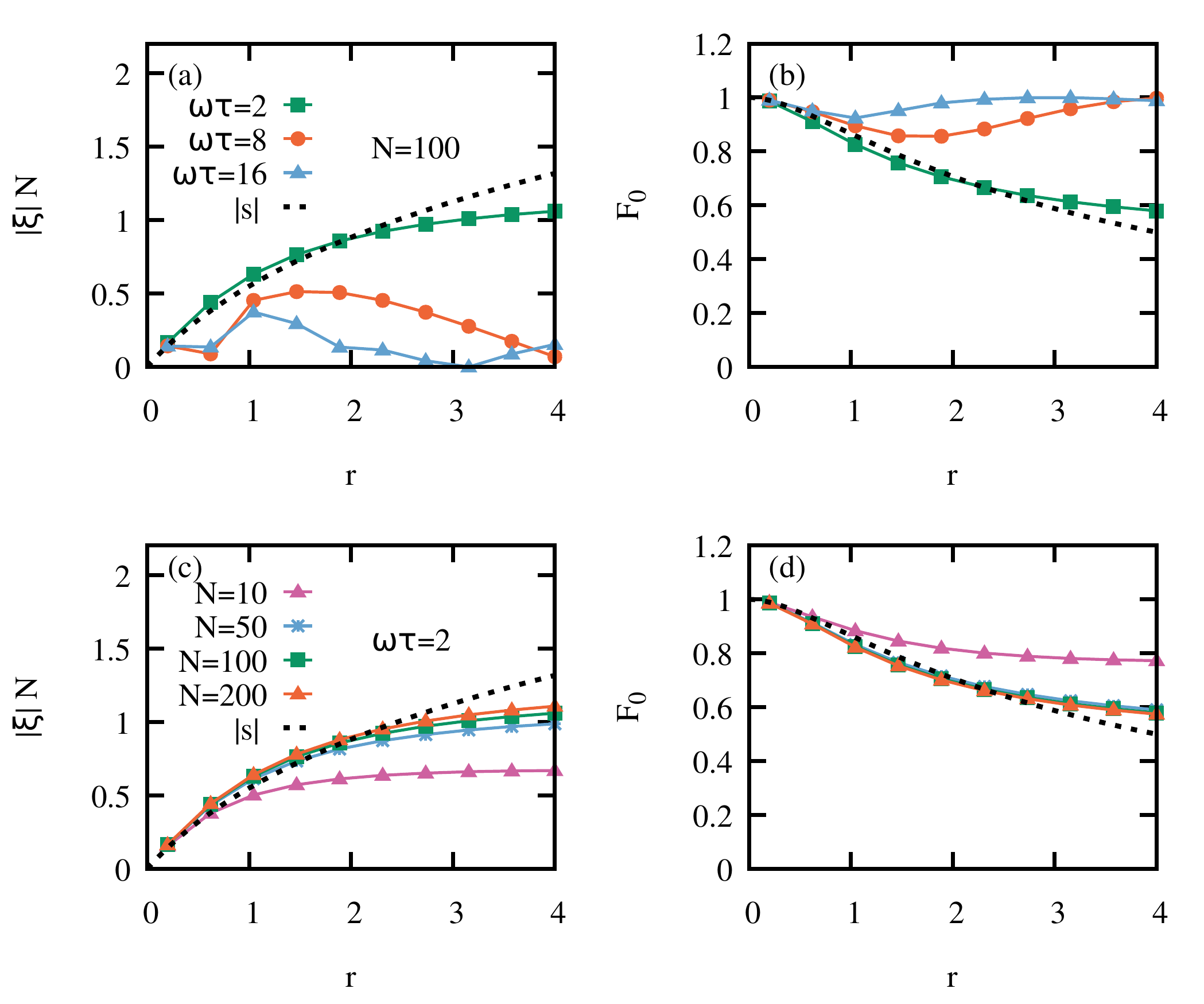}
\caption{\small{Generated spin squeezed states using a cycle toward the critical point $g_c$ using a protocol $g(t)$ characterized by the nonlinear exponent $r$. Top panels (a) and (b) show the resulting squeezing parameter $|\xi| N$ and the ground-state fidelity $F_0$ for a fixed number of spins $N=100$ versus the nonlinear exponent $r$ and varying the protocol duration $\tau$. Dashed lines are the theoretical predicted values in the thermodynamic limit where $z\nu=1/2$ (cf. Eqs.~\eqref{eq:xiN} and~\eqref{eq:f0}). As illustrated in (a), the slower the cycle is performed the more adiabatic it becomes due to a non-zero energy gap at $g_c$, i.e. due to finite-size effects. For $\omega\tau=16$ the protocol is not able to generate non-adiabatic excitations and the ground state fidelity stay close to 1 (b). Large values for $r$ imply larger squeezing, but finite size effects are more prominent (see bottom panels). Bottom panels (c) and (d): Similar to top panels but varying the system size $N$ to better illustrate the finite-size effects.}}
\label{figLMG_SM}
  \end{figure}

Here we provide more details for the generation of spin squeezed states in the Lipkin-Meshkov-Glick (LMG) model for $\kappa=0$. As commented in the main text, we numerically solve the Schr{\"o}dinger equation $d/dt \ket{\psi(t)}=-iH_{\rm LMG}(t)\ket{\psi(t)}$ where $H_{\rm LMG}(t)$ is given in Eq.~(7) of the main text. In this way, we obtain the final state after one cycle $\ket{\psi(2\tau)}$. In order to verify our theoretical predictions, we find $|\xi|$ and $\beta$, such that $\xi=|\xi|e^{i\beta}$, that maximize the fidelity $F_\xi$, 
\begin{align}
    F_\xi=|\langle\psi(2\tau)|\xi \rangle|^2,
\end{align}
where $\ket{\xi}=S_{\rm spin}(\xi)\ket{J,m_J=J}$ is a spin squeezed state. As shown in Fig. 3, the found parameters closely follow the theoretical prediction $|\xi|=|s|/N$, i.e.
\begin{align}\label{eq:xiN}
    |\xi|N = |s|={\rm arcosh}\left(\csc\left(\frac{\pi}{2+2z\nu r}\right)\right).
\end{align}
In addition, here we provide further information and results regarding the generated spin squeezed states (cf. Fig.~\ref{figLMG_SM}). As emphasized in the main text, the total duration $\tau$ cannot be arbitrarily large to avoid the adiabatic condition, that is, $\epsilon(g_c)\tau\lesssim 1$ where $\epsilon(g_c)$ denotes the energy gap of $H_{\rm LMG}$ at the critical point $g_c=1$, while at the same time $\omega\tau\gtrsim 1$ which excludes sudden-like cycles. The impact of choosing larger $\omega\tau$, exceeding this constrain is illustrated in Fig.~\ref{figLMG_SM}(a) and (b). For $N=100$, the energy gap $\epsilon(g_c)\approx 0.5\omega$, and therefore $\omega\tau=2$ fulfills both criteria, while  $\omega\tau=8$ and $16$ do not. Cycles performed with the latter $\omega\tau$ values result in the suppression of non-adiabatic excitations, and thus a significant deviation from the expected squeezing. The ground-state fidelity is another good indicator, $F_0=|\langle J,m_J=J|\psi(2\tau)\rangle|^2$. If the cycle is performed adiabatically, $F_0=1$, while if $\ket{\psi(2\tau)}=\ket{\xi}$ with $|\xi|=|s|/N$, it follows
\begin{align}\label{eq:f0}
    F_0=\sin\left(\frac{\pi}{2+2z\nu r}\right).
\end{align}
Note that the previous expression holds in the thermodynamic limit $N\rightarrow \infty$. 
As the theoretical prediction has been derived in the $N\rightarrow\infty$, increasing $N$ leads to a better agreement with Eq.~\eqref{eq:xiN} and~\eqref{eq:f0}. This is plotted in Fig.~\ref{figLMG_SM}(c) and (d). Finally, we note that all the generated states in Fig.~\ref{figLMG_SM}(c) and (d) are entangled and provide a metrological gain since $1>\chi^2_{\rm min}\gtrsim 0.1$.

\begin{figure}
    \centering
    \includegraphics[width=0.6\linewidth,angle=0]{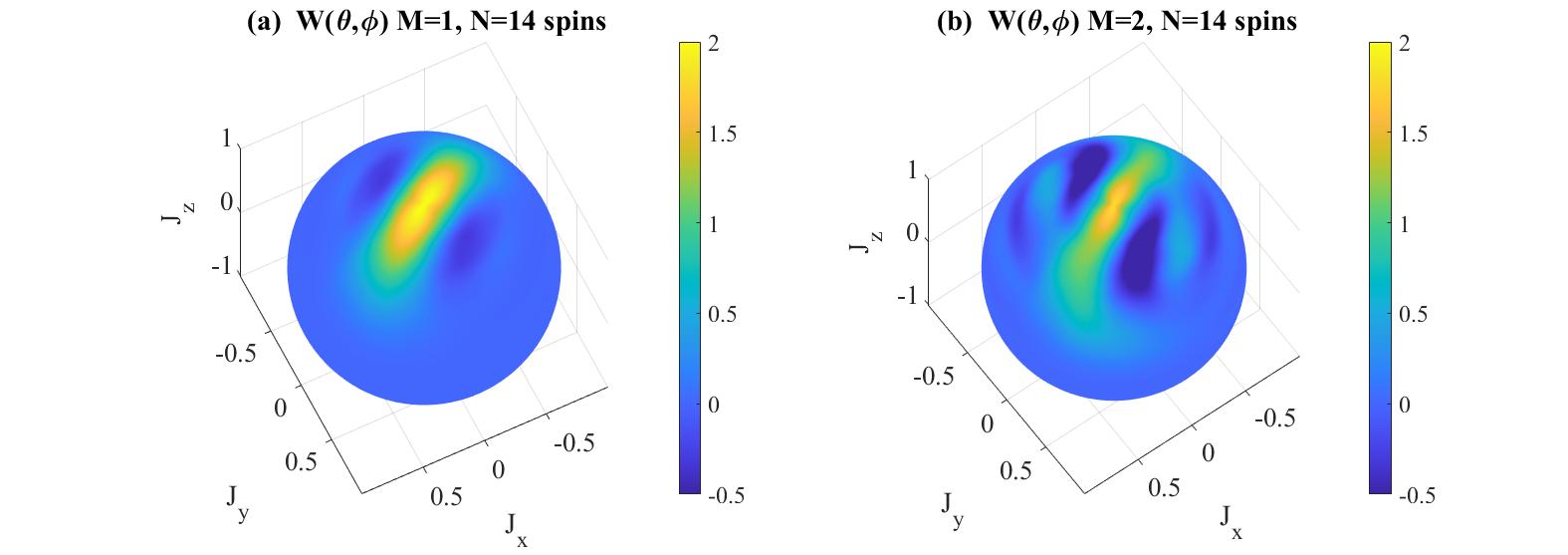}
    \caption{\small{Wigner quasiprobability distribution $W(\theta,\phi)$ for the resulting spin states in the LMG for $N=14$ spins, after $M=1$ (a) and $M=2$ cycles (b) with $\omega\tau=2$ and $r=2$, as shown in Fig. 3(d) in the main text. }}
    \label{figLMG_Wigners}
\end{figure}

We follow Ref.~\cite{Dowling:94SM} to obtain the Wigner quasi-probability distribution $\mathcal{W}(\phi,\theta)$ for an ensemble of $N$ spin-$\frac{1}{2}$ particles. In particular, $\mathcal{W}(\phi,\theta)$ can be written as
\begin{align}
    \mathcal{W}(\phi,\theta)=\sum_{k=0}^{2J}\sum_{q=-k}^{k}Y_{kq}(\theta,\phi)\rho_{kq}, \quad {\rm with} \quad \rho_{kq}={\rm Tr}[\rho T_{kq}^\dagger],
\end{align}
where the operator $T_{kq}$ is defined as follows
\begin{align}
    T_{kq}=\sum_{m=-J}^J \sum_{m'=-J}^J (-1)^{J-m}\sqrt{2k+1} \begin{pmatrix}
J & k & J\\
-m & q & m'
\end{pmatrix} \ket{J, m}\bra{J,m'},
\end{align}
and $Y_{kq}(\theta,\phi)$ denotes the usual spherical harmonics,
\begin{align}
    Y_{kq}(\theta,\phi)=(-1)^q \sqrt{\frac{(2k+1)}{4\pi}\frac{(k-q)!}{(k+q)!}} P_{kq}(\cos\theta) e^{iq\phi},
\end{align}
with $P_{kq}(x)$ the Legendre polynomials. In this manner, the Wigner function is normalized according to
\begin{align}
    \int d\Omega \mathcal{W}(\phi,\theta)=\left(\frac{4J+1}{4\pi} \right)^{-1/2}.
\end{align}
See Ref.~\cite{Dowling:94SM} for further details on the Wigner $\mathcal{W}(\phi,\theta)$ for spin states. In Fig.~\ref{figLMG_Wigners} we show the resulting Wigner quasiprobability distributions for $M=1$ and $M=2$ cycles, to illustrate the increase of squeezing and metrological gain obtained by repeating cycles. Note that the case for $M=3$ corresponds to Fig. 3(d) in the main text.

\end{document}